\documentclass{article}

\usepackage{arxiv}

\usepackage{comment}

\usepackage[utf8]{inputenc} 
\usepackage[T1]{fontenc}    
\usepackage{hyperref}       
\usepackage{url}            
\usepackage{booktabs}       
\usepackage{amsfonts}       
\usepackage{nicefrac}       
\usepackage{microtype}      
\usepackage{xcolor}         

\usepackage{graphicx}
\usepackage{subcaption}
\usepackage{amsmath}
\usepackage{wrapfig}

\title{Thinking Outside the Box: Orthogonal Approach to Equalizing Protected Attributes}

\author{%
  Jiahui Liu\textsuperscript{1},
  Xiaohao Cai\textsuperscript{1},
  Mahesan Niranjan\textsuperscript{1}\\
  \textsuperscript{1}University of Southampton, Southampton, UK\\
  \texttt{\{jl4f19, x.cai, mn\}@soton.ac.uk}
}

\begin{document}

\maketitle

\begin{abstract}
There is growing concern that the potential of black box AI may exacerbate health-related disparities and biases such as gender and ethnicity in clinical decision-making. Biased decisions can arise from data availability and collection processes, as well as from the underlying confounding effects of the protected attributes themselves. This work proposes a machine learning-based orthogonal approach aiming to analyze and suppress the effect of the confounder through discriminant dimensionality reduction and orthogonalization of the protected attributes against the primary attribute information. By doing so, the impact of the protected attributes on disease diagnosis can be realized, undesirable feature correlations can be mitigated, and the model prediction performance can be enhanced.
\end{abstract}

\section{Introduction}
Machine/deep learning (ML) has earned significant attention in the medical field, offering state-of-the-art solutions in enhancing disease diagnosis and treatment management and broadening healthcare accessibility. As AI systems gain traction in medical imaging diagnosis, there is a growing awareness about the imperative need for fairness guarantee in the systems' prediction and the investigation of latent biases which may emerge in intricate real-world scenarios \cite{chen2021ethical, lee2022towards}. Unfortunately, AI models often inadvertently encode sensitive attributes (such as race and gender) when processing medical images, thereby influencing their discriminatory behaviour \cite{glocker2021algorithmic, zhang2022improving, dankwa2022artificial}. This issue becomes particularly noticeable when models are trained on data sourced from external repositories but are evaluated on data from internal ones. Therefore, while the diagnosis remains consistent across datasets, differences in protected attributes can lead to suboptimal model performance on the internal datasets \cite{degrave2021ai}. 
This perspective holds merit as bias is inherently human and deeply ingrained in our society, and is therefore challenging to be eliminated entirely. The urgent need lies in constructing AI/ML systems in healthcare that are equitable, transparent and impartial.

Efforts are underway to address the above-mentioned issue across various diagnostic domains. The work in \cite{gichoya2022ai} undertook a study that goes beyond predicting a patient's age and biological sex \cite{yi2021radiology}, highlighting the remarkable capability of deep neural networks in accurately discerning a patient's racial identity from medical scans including chest X-rays. It specifically demonstrated the potential of models to incorporate protected attributes from images into their predictions. These findings also hold significant implications, i.e., AI/ML models could potentially amplify health disparities. The work in \cite{raza2022machine} tackled this issue with a numerical hyperglycemia dataset, emphasizing the importance of eliminating biases during data collection to prevent their negative impact on model inference. 
Methods based on disentanglement achieved separation by reducing the mutual information between target and sensitive attribute representations \cite{locatello2019challenging, mahapatra2022unsupervised}. 

Due to the `black box' nature of the majority of AI/ML models, it is challenging to discern the specific information being utilized for models' prediction. In particular, the feature representations generated in the penultimate layer of neural networks lack semantic information, making it difficult to ascertain the potential use of implicit/protected attributes. In this paper, our key contribution is leveraging discriminant analysis to enforce orthogonality between protected attributes and the primary target, thus facilitating disentanglement and enhancing model performance.

\section{Method}
\paragraph{Preliminary.} We form a data matrix $\boldsymbol{Y} = (\boldsymbol{y}_1, \boldsymbol{y}_2, \cdots, \boldsymbol{y}_N)^\top \in \mathbb{R}^{N\times M}$ by $N$ number of image samples  $\boldsymbol{y}_i = (y_{i1}, y_{i2}, \cdots, y_{iM})^\top \in \mathbb{R}^M$, where $M$ is the number of features of every sample. These $N$ samples are associated with $C$ distinct classes, namely $\boldsymbol{\Lambda}_j$ representing their primary labels for class $j$, and their cardinality $|\boldsymbol{\Lambda}_j| = N_j$, $j\in [C]$, where $[C]$ denotes the set containing the first $C$ natural numbers. Let $\bar{\boldsymbol{y}}$ and $\bar{\boldsymbol{y}}_j$ respectively be the mean of the whole samples and the samples in class $j$, i.e., $\bar{\boldsymbol{y}} = \frac{1}{N} \sum_{i=1}^N \boldsymbol{y}_i$, $\bar{\boldsymbol{y}}_j = \frac{1}{N_j} \sum_{\boldsymbol{y} \in \boldsymbol{\Lambda}_j} \boldsymbol{y}$, $j\in [C]$.
Let $\boldsymbol{S}_{\rm B}$ and $\boldsymbol{S}_{\rm W}$ denote the inter- and intra-class scatters, respectively, i.e.,
$
\boldsymbol{S}_{\rm B}  =\sum_{j=1}^C(\bar{\boldsymbol{y}}_j - \bar{\boldsymbol{y}})(\bar{\boldsymbol{y}}_j - \bar{\boldsymbol{y}})^{\top},  
\boldsymbol{S}_{\rm W} =\sum_{j=1}^C  \boldsymbol{S}^j_{\rm W},
$
where $\boldsymbol{S}^j_{\rm W}= \sum_{k=1}^{N_{j}}(\boldsymbol{y}_k^j-\bar{\boldsymbol{y}}_j) (\boldsymbol{y}_k^j-\bar{\boldsymbol{y}}_j)^{\top}, j\in [C]$.
Additionally, those $N$ samples are annotated with labels indicating their protected attributes, creating $D$ distinct classes, namely $\boldsymbol{\Delta}_k$ representing their protected attributes for class $k \in [D]$. Analogously, the inter- and intra-class scatters
can be formed according to $\boldsymbol{\Delta}_k$, say 
$\boldsymbol{S}_{\rm B}^\dagger$ and $\boldsymbol{S}_{\rm W}^\dagger$. For simplicity, we consider the binary classification problem, i.e., $C=D=2$. In this case, $\boldsymbol{S}_{\rm B}$ can be simplified as $\boldsymbol{s}_{\rm b} \boldsymbol{s}_{\rm b}^\top$, 
where $\boldsymbol{s}_{\rm b} = \bar{\boldsymbol{y}}_1 - \bar{\boldsymbol{y}}_2$.

\paragraph{Methodology.} The rationale of our method is briefly described below. Step 1, a pre-trained deep neural network (e.g. ResNet18 in our case) originally designed to solve large natural image classification problems is selected and then used here to extract features of medical images from the given medical imaging datasets. Step 2, an exquisite discriminant analysis technique is designed to project the extracted features into a lower-dimensional space by forcing the protected attributes in the given medical imaging data orthogonal to the primary target (details see below). Step 3, a classifier like SVM is applied for the final prediction with parameters fine-tuned using Bayesian optimization.

\begin{wrapfigure}{r}{0.6\linewidth}
\vspace{-0.23in}
\centering
\includegraphics[width=3.3in, height=1.3in]{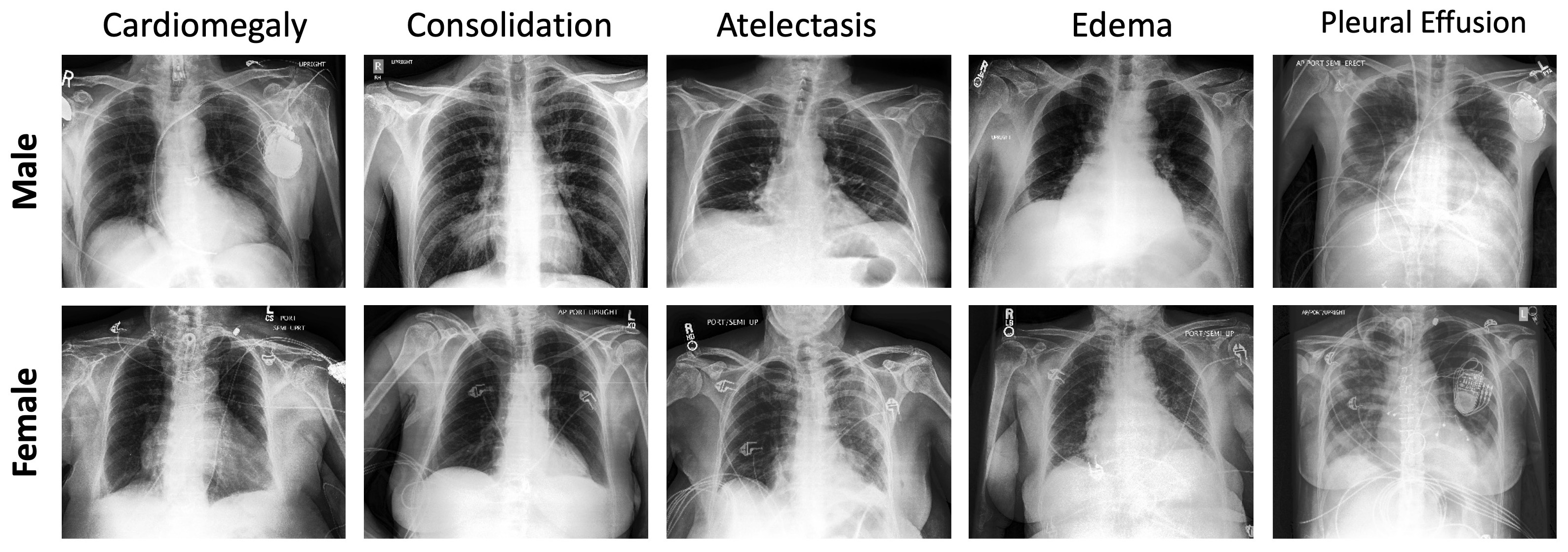}
\caption{{\small Some X-ray scans from the CheXpert data. }}
\label{Gender}
\vspace{-0.15in}
\end{wrapfigure}

The above Step 2 plays the key role in our method. For binary classification problem, we are looking for two discriminant directions, say $\boldsymbol{d}_1, \boldsymbol{d}_2 \in {\mathbb{R}}^{M}$, which are orthogonal and, more importantly, exploiting the image labels (i.e., the primary labels) and the protected attributes, respectively. 

We ask for the first direction $\boldsymbol{d}_1$ to maximize the Fisher criterion, i.e., ${\cal R}(\boldsymbol{d}) = \boldsymbol{d}^{\top} \boldsymbol{S}_{\rm B} \boldsymbol{d} / {\boldsymbol{d}^{\top} \boldsymbol{S}_{\rm W} \boldsymbol{d}}$.
Then we have $\boldsymbol{d}_{1} = \alpha_1 {\boldsymbol{S}_{\rm W}}^{-1}\boldsymbol{s}_{\rm b}$, where $\alpha_1$ is the normalizing constant such that $\|\boldsymbol{d}_{1}\|_2 = 1$ (i.e., $\alpha_{1}^{2}=(\boldsymbol{s}_{\rm b}^{\top}[\boldsymbol{S}_{\rm W}^{-1}]^{2} \boldsymbol{s}_{\rm b})^{-1}$); see e.g. \cite{foley1975optimal,liu2023go} for the derivation. For $ \boldsymbol{d}_{2}$, since we need it to involve the information of the protected attributes $\boldsymbol{\Delta}_k$, we require it to maximize another Fisher criterion, i.e., ${\cal R}^\dagger(\boldsymbol{d}) = \boldsymbol{d}^{\top} \boldsymbol{S}^\dagger_{\rm B} \boldsymbol{d} / {\boldsymbol{d}^{\top} \boldsymbol{S}^\dagger_{\rm W} \boldsymbol{d}}$, with the orthogonal constraint $\boldsymbol{d}_{2}\perp \boldsymbol{d}_{1}$. This optimisation problem of finding $\boldsymbol{d}_{2}$ is equivalent to find the eigenvector corresponding to the largest eigenvalue of the generalized eigenvalue problem, i.e., 
\begin{equation} \label{eqn:go-lda-k1}
\big(\boldsymbol{S}^\dagger_{\rm B}-\boldsymbol{k}_{1}\big) \boldsymbol{d} = {\mu} \boldsymbol{S}^\dagger_{\rm W} \boldsymbol{d}, \quad \textrm{with} \ \
\boldsymbol{k}_{1} = 
{\boldsymbol{d}_{1}^{\top} [\boldsymbol{S}_{\rm W}^\dagger]^{-1}  \boldsymbol{S}^\dagger_{\rm B} \boldsymbol{d}_{1}} / \big({\boldsymbol{d}_{1}^{\top} [\boldsymbol{S}_{\rm W}^\dagger]^{-1} \boldsymbol{d}_{1}}\big).
\end{equation}
The derivation in Equation \eqref{eqn:go-lda-k1} can be found in \cite{liu2023go}. After finding the orthogonal discriminant directions $\boldsymbol{d}_1$ and $\boldsymbol{d}_2$, they can subsequently be applied to reduce the dimension from the original feature space to the desired subspace, effectively decoupling the primary task from the subtask.

\section{Experimental Results}
\label{dataset}
We showcase the performance of our method on the CheXpert dataset\footnote{https://stanfordmlgroup.github.io/competitions/chexpert/}, which encompasses five distinct diseases, i.e., cardiomegaly, consolidation, atelectasis, edema and pleural effusion. Note that each disease will form a different binary classification problem. The protected attributes are gender, i.e., male or female. The X-ray scans regarding different genders under different diseases are showcased in Figure \ref{Gender}. For ease of reference, we refer the way in our method by dropping the dimensionality reduction step (i.e., Step 2) to {\it baseline} for comparison.


\begin{figure}[ht]
\vspace{-0.11in}
\centering
\includegraphics[trim={{.0\linewidth} {.0\linewidth} {.0\linewidth} {.29\linewidth}}, clip, width=5.5in, height=2.4in]{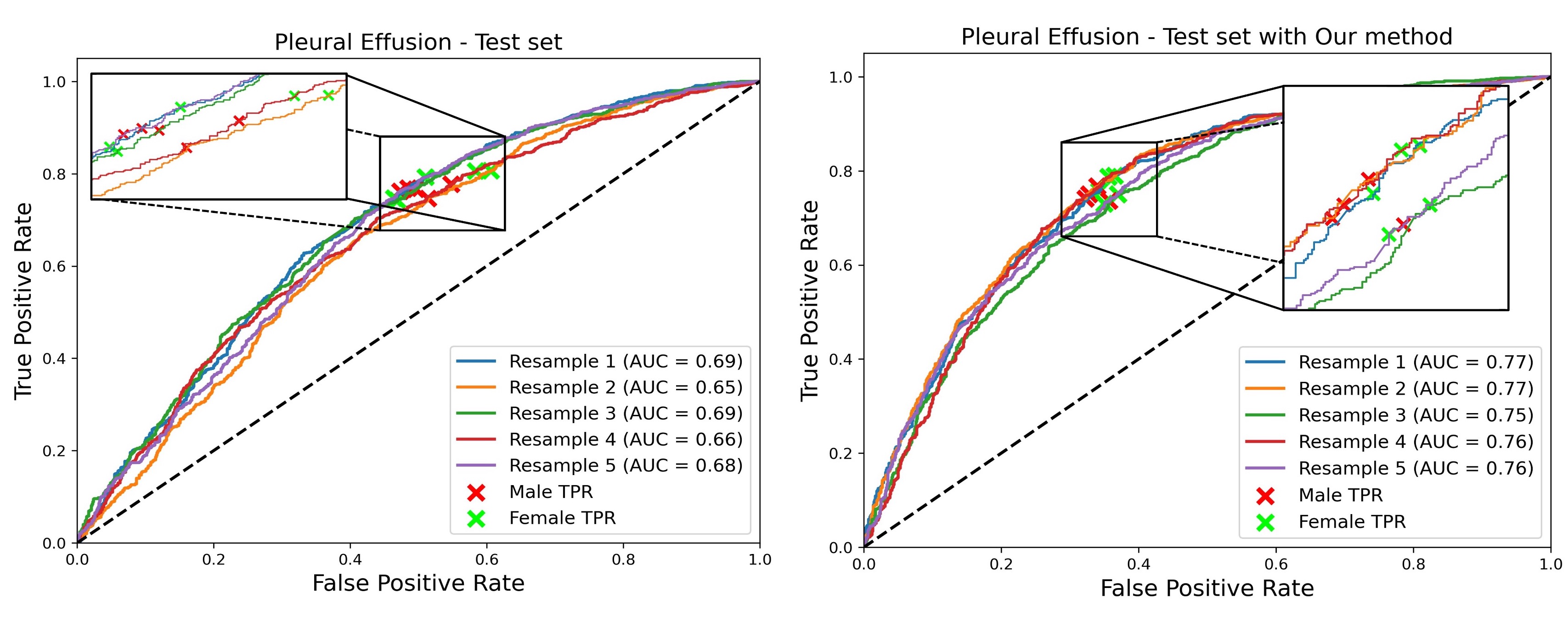}
\vspace{-0.25in}
\caption{Comparison between the baseline ({\it left} panel) and our method ({\it right} panel) regarding the gender effect in the detection of the disease `pleural effusion' (detailed description in the main text).}
\label{PleuralEffusion}
\vspace{-0.13in}
\end{figure}

Figure \ref{PleuralEffusion} presents the investigation of the gender effect in the detection of the disease `pleural effusion' across five distinct resampled test datasets. The left panel (baseline) shows that the area under the curve (AUC) values and the shapes of the receiver operating characteristic (ROC) curves remain consistent across different resampled datasets; however, there exists a variation in the true positive rate (TPR) between male and female  patients across all these different resampled datasets. After involving the protected attributes (i.e., gender) by our method, it is evident that the impact of gender has significantly diminished, see the right panel. It is worth highlighting that our method also improves the average AUC values by 8.8\%, demonstrating its superior prediction ability in this problem.

\begin{wrapfigure}{r}{0.66\linewidth}
\vspace{-0.23in}
\centering
\includegraphics[width=3.6in, height=1.6in]{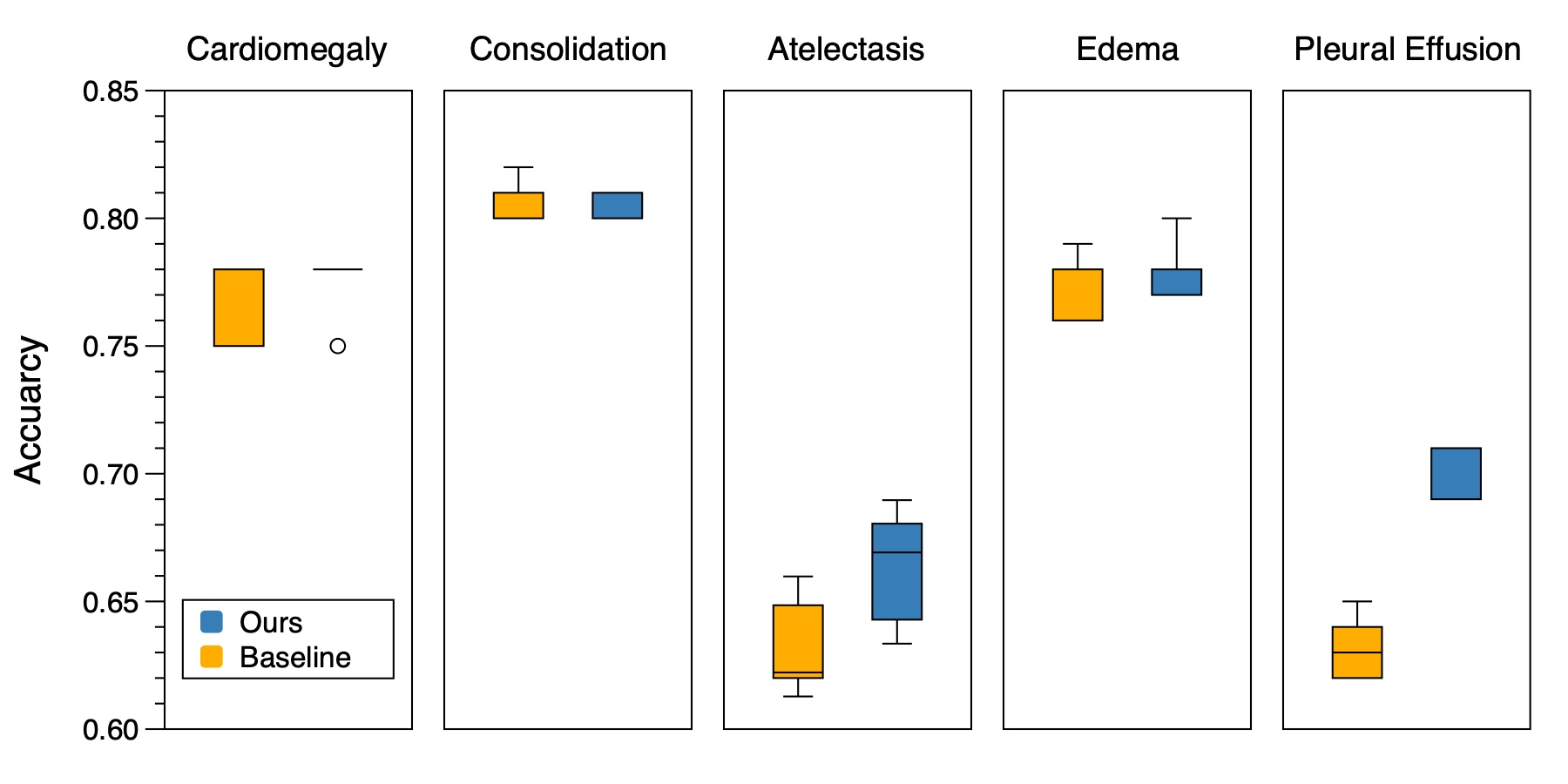}
\vspace{-0.2in}
\caption{Comparison regarding every disease in CheXpert. } \label{Five_disease}
\vspace{-0.1in}
\end{wrapfigure}

Figure \ref{Five_disease} illustrates the impact of the gender factor on each of the five different diseases in the CheXpert dataset. It firstly indicates that the gender attribute does not consistently exert great impact across all diseases. It is also evident that, in the case of certain disease like `pleural effusion', the influence of gender is paramount. These findings are of great importance in clinical diagnosis for example. It could assist doctors to reach more accurate decision in a fairer manner, which can instil greater confidence when incorporating external data into hospital practices.

In addition to the ability of identifying the impact of the gender attribute on each disease, Figure~\ref{Five_disease} again shows that our method can also enhance the prediction accuracy by a large margin through exploiting the protected attributes to correct the intrinsic biases within the data.


\section{Conclusion and Future Research}
In this work, we proposed a novel method by exploiting exquisite orthogonal discriminant directions and pre-trained deep neural networks to effectively analyze implicit biases raised by protected attributes and enhance prediction performance in medical imaging.
It may not perform optimally with extremely large datasets like large foundation models. However, it is attributed to the unique feature of medical images, where acquiring a substantial quantity of data is challenging compared to natural images. Future work will primarily focus on its application in mitigating implicit biases in medical imaging with a wider range of data and various protected attributes.

\newpage

\bibliographystyle{plain}
\bibliography{refs}

\newpage



\end{document}